\documentclass[aps,prd,twocolumn,nofootinbib,preprintnumbers]{revtex4}
\usepackage[hypertex]{hyperref}
\usepackage{graphicx}
\usepackage{color}

\begin{document}



\title{A Little Twin Higgs Model}

\author{Hock-Seng Goh} \author{Christopher A. Krenke}
\affiliation{Department of Physics, University of Arizona, Tucson,
AZ 85721}


\begin{abstract}

We present a twin Higgs model based on left-right symmetry with a tree level quartic.
This is made possible by extending the symmetry of the model to include two $Z_2$
parities, each of which is sufficient to protect the Higgs from getting a
quadratically divergent mass squared. Although both parities are broken explicitly,
the symmetries that protect the Higgs from getting a quadratically divergent mass are
broken only collectively. The quadratic divergences of the Higgs mass are thus still
protected at one loop. We find that the fine tuning in this model is reduced
substantially compared to the original left-right twin Higgs model. This mechanism can
also be applied to the mirror twin Higgs model to get a significant reduction of the
fine tuning, while keeping the mirror photon massless.

\end{abstract}

 \maketitle


\section{Introduction}
The standard model (SM) is so far the most successful theory that
describes physics at energies below the TeV scale. Its
predictions are consistent with all precision electroweak
measurements. However, the model is unsatisfactory since the Higgs
field, which plays a crucial role in electroweak symmetry
breaking, receives quadratically divergent radiative corrections
to its mass and thus destabilizes the electroweak scale. Hence, it
is unnatural to treat the SM as an effective theory with a cutoff scale much
higher than a TeV. On the other hand, the cutoffs of nonrenormalizable operators
that contribute to precision electroweak measurements are required by experiment to
be greater than 5-10 TeV. Such a high cutoff tends to destabilize
the electroweak scale and leads to a fine tuning of a few
\%. This problem is known as the little hierarchy problem or the
LEP paradox \cite{LEPparadox}.

The idea that the Higgs is a pseudo-Nambu-Goldstone boson (PNGB)
corresponding to a spontaneously broken global symmetry was
proposed in refs. \cite{GP,KG}. Since the mass of a PNGB tends to
be lighter than the UV scale, this idea explains why the Higgs is
light. However, using this idea to solve the little hierarchy
problem is not quite straightforward. A PNGB Higgs by itself is
not sufficient since the global symmetry is, by definition, not
exact and the couplings that break the global symmetry will still
generate a quadratically divergent mass to the Higgs. Thus, the
situation is no better than that in the standard model and more
structure is needed.  The extra structure required to achieve
naturalness is the main challenge for model building. One
successful mechanism along this line is known as the little Higgs
\cite{Little1,Little2}. In this class of models, the Higgs mass is
protected by two separate global symmetries and every term in the
Lagrangian breaks at most one of them. In order to break both
global symmetries, radiative corrections to the mass have to
involve at least two such terms and thus, quadratic divergences
are postponed to two loops. This little Higgs mechanism is also
known as collective symmetry breaking. To achieve a certain level
of naturalness, a special operator is also introduced to provide a
tree level quartic without generating a tree level mass to the
Higgs.

Another mechanism that has been shown to solve the little
hierarchy problem is the twin Higgs
\cite{chacko,twinhiggsmirror,nomura,lrtwin} (see also
\cite{susytwinhiggs,UEDtwinhiggs,foot}). The twin Higgs mechanism
is quite different from that of the little Higgs. In twin Higgs
models, the Higgs mass is protected by a discrete $Z_2$, or twin,
symmetry instead of multiple global symmetries. The exact twin
symmetry guarantees that all gauge invariant dimensionful terms
have, up to all orders in perturbation theory, a form which is
invariant under a global SU(4) symmetry. The mass of the PNGB
Higgs is then protected from receiving quadratically divergent
contributions. It was shown that this mechanism alleviates the
little hierarchy problem to about the 10\% level for the cut off
scale $\Lambda=10$ TeV without introducing a tree level quartic.

In this class of models where the quadratic divergences are
naturally suppressed, one would expect less fine-tuning if the
quartic coupling of Higgs $\lambda$ is large. In the original twin
Higgs models, both the squared mass and the quartic for the SM
Higgs come from the one-loop Colemann-Weinberg (CW) potential
\cite{CW}. The quartic coupling is thus not a free parameter and
loop suppressed. In order to improve the naturalness, one should
try to find a tree level operator that will give the PNGB a
quartic coupling without giving it  a tree level mass term. In
order to not upset the cancellation of radiative corrections, the
tree level operator one introduces must preserve the twin parity.
To summarize, in order to improve the fine tuning, the following
criteria must be satisfied.

\begin{itemize}
    \item A tree level operator that generates a quartic
    for the SM Higgs, but not a mass.
    \item This operator must preserve the discrete symmetry that protects the
    Higgs mass.
    \item Reduce as much as possible the mass squared that arise at
    loop level. For example, reduce top contribution by making the top Yukawa interaction SU(4) invariant.
\end{itemize}

One very simple operator which satisfies the first criterion has been constructed and
is used in the twin Higgs model \cite{nomura}.  The basic idea is a mismatched
alignment of two vevs. It was shown in ref. \cite{nomura} that the mirror twin Higgs
model \cite{chacko} improves when this type of tree level quartic is added. The
mismatched alignment of the vevs necessarily breaks the mirror $SU(2)\times U(1)$
gauge symmetry to nothing and so the mirror photon becomes massive. Because of this
feature, the mechanism seems difficult to implement in the left-right twin Higgs model
\cite{lrtwin} since the mismatched vev alignment would break $U(1)_{EM}$ and the SM
photon would become massive.

However, there is actually more than one type of parity which can
be identified as a twin parity, i.e. the original twin parity,
known also as ${\bf P}$, and charge conjugation, ${\bf C}
$\cite{twolrs}. Under these parities, scalar and Dirac fermion in
the left-right model transform as
\begin{eqnarray}
    {\bf P} : \left\{%
    \begin{array}{rcl}
      H_L & \rightarrow & H_R \\
      Q_L & \rightarrow & Q_R \\
    \end{array}\right\}%
\end{eqnarray}
\begin{eqnarray}
    {\bf C} : \left\{%
    \begin{array}{rcl}
      H_L & \rightarrow & H_R^* \\
      Q_L & \rightarrow & C\bar{Q}^T_R \\
    \end{array}\right\}%
\end{eqnarray}
In this paper, we show that by using this fact and the idea of
collective symmetry breaking, a new type of quartic operator can
be constructed. This new quartic has all the properties we
mentioned above, but does not break $U(1)_{EM}$. Most importantly,
it preserves one of the parities that will maintain the
cancellation of quadratic divergences to one loop. The quadratic
divergences are no longer protected to all orders in perturbation
theory as in the original twin Higgs model. However, cancellation
to one loop is sufficient to address the little hierarchy problem.

The paper is organized as follows:  In section II, we review the
twin Higgs mechanism and the left-right twin Higgs model. We then
explore the possibilities in introducing a tree level quartic and
extend the top sector, making it SU(4) invariant. In section III,
we analyze the radiative corrections and electroweak symmetry
breaking.  We then apply the same mechanism to the mirror model
and reanalyze its naturalness in section IV. In section V, some
phenomenology is discussed and our results summarized.

\section{Construction of the Model}

The scalar field $H$ in twin Higgs models is in the fundamental
representation of a U(4) global symmetry. After acquiring a vev,
$\langle H \rangle=(0,0,0,f)$, U(4) is broken to U(3), which
yields 7 Goldstone bosons including the standard model (SM) Higgs
doublet $h=(h_1,h_2)$. The global symmetry is explicitly broken by
gauging only a subgroup $SU(2)_A\times SU(2)_B$ (we ignore $U(1)$
factors here since they are not relevant to present discussion).
Under this gauge symmetry, H can be represented by $H=(H_A,H_B)$
where $H_{A,B}$ transform as doublets of $SU(2)_{A,B}$. Since the
global symmetry is broken explicitly by the gauge couplings and
the breaking is `hard', masses of the Goldstone bosons will be
radiatively generated and be quadratically divergent. However, by
imposing the discrete symmetry (twin parity) that interchanges the
two gauged $SU(2)$ symmetries , the quadratic divergences cancel.
The simplest way to understand this is the following. First write
down the most general gauge invariant mass terms for the linear
fields $H_A$ and $H_B$
\begin{equation}
    \alpha_A H_A^{\dagger}H_A+\alpha_B H_B^{\dagger}H_B,
\end{equation}
where $\alpha_{A,B}$ are not required to be related by the gauge symmetry. After
imposing the twin symmetry on all the interactions, however, $\alpha_A$ is forced to
be equal to $\alpha_B$ and so the form given above is invariant under the global U(4)
transformation. Therefore, this term, which is quadratically divergent, does not
contribute to potential of the Goldstone bosons. Higher order terms, the quartic term
$(|H_A|^4+|H_B|^4)$ for example, can contribute even though they preserve the twin
symmetry since twin symmetry does not require these terms to have a U(4) invariant
form. These contributions can have at most logarithmic divergences and so are under
theoretical control. Additional interactions such as Yukawa couplings can be added to
the theory consistent with the discrete twin symmetry, and the argument above shows
that they do not lead to quadratic divergences.

The fine-tuning in twin Higgs theories can be further reduced if
there are terms in the Lagrangian which respect the twin symmetry
and contribute to the quartic self-coupling of the light
pseudo-Goldstone Higgs but not to its mass. In the case of the
model discussed above, with a single Higgs field $H$, there are no
such operators consistent with the symmetries of the theory.
However, such terms can be written down in theories with more than
one set of Higgs fields. We consider the theory with an extra
scalar field $\hat{H}$, which has its vev residing in a different
direction, $\langle\hat{H}\rangle=(0,0,\hat{f},0)$\cite{nomura}.
After the global U(4) symmetry is spontaneously broken by $f$ and
$\hat{f}$, and the massive radial modes are integrated out, we can
write down a non-linear sigma model which contains the interactions
of the light degrees of freedom. The light fields of the non-linear sigma
model can be parametrized as

\begin{eqnarray}\label{H_expand}
    H = \left(
\begin{array}{c}
  h_1 \\
  h_2 \\
  C \\
  f+i\phi-\frac{h^{\dagger}h}{2f} \\
\end{array}%
\right)+\cdots\nonumber\\
\hat{H}=\left(%
\begin{array}{c}
  \hat{h}_1 \\
  \hat{h}_2 \\
  \hat{f}+i\hat{\phi}-\frac{\hat{h}^{\dagger}\hat{h}}{2\hat{f}} \\
  \hat{C} \\
\end{array}
\right)+\cdots
\end{eqnarray}
Notice that a quartic term like $|H|^4$ would give a mass term to
the Goldstone boson $h$ because $H$ contains a component
$\sim(f-h^2/2f+...)$. The quartic operator $|H|^4$ therefore
contains a term like $(f-h^2/2f+...)^4$, which gives a mass term
for $h$. This is why the second Higgs field $\hat{H}$ is required.
With the mismatched alignment of vevs as in eq.~(\ref{H_expand}),
the operator $|H^{\dagger}\hat{H}|^2$ gives mass only to $C$ and
$\hat{C}$, and gives rise to a quartic term for $h$ and $\hat{h}$
without a corresponding mass term.

The above discussion is general for twin Higgs models. The
phenomenological consequences of the additional vev $\hat{f}$,
however, depend on the model's U(1) structure. In the mirror twin
Higgs model, the gauged subgroup of global U(4) is $SU(2)_A\times
U(1)_A\times SU(2)_B\times U(1)_B$. Two identical electroweak
gauge symmetries are introduced to two sectors of the model.
Sector A is identified with the standard model and sector B is a
mirror world of the standard model. An extra scalar multiplet
$\hat{H}=(\hat{H}_A,\hat{H}_B)$ is added to the model in order to
implement the above mechanism \cite{nomura}. $H_B$ and $\hat{H}_B$
are both singlets under $SU(2)_A\times U(1)_A$ and have the same
nontrivial charge under $SU(2)_B\times U(1)_B$. The mismatched
vevs thus break the mirror $SU(2)_B\times U(1)_B$ gauge symmetry
to nothing but preserve the entire standard model $SU(2)_A\times
U(1)_A$. The mirror photon therefore becomes massive, in contrast
to the case in the original mirror twin Higgs model where the
mirror photon remains massless after U(4) symmetry is broken.

In the left-right twin Higgs (LRTH) model, the gauged subgroup is
that of the left-right model: $SU(2)_L\times SU(2)_R\times
U(1)_{B-L}\times {\bf P}$\cite{leftright}. There are two Higgs
fields, $H=(H_L,H_R)$ and $\hat{H}=(\hat{H}_L,\hat{H}_R)$, both of
which transform as a fundamental representation under the SU(4)
global symmetry. Under the gauge symmetry, these scalars transform
as
\begin{equation}\label{Hcharge}
    H_L {\rm \ and\ }\hat{H}_L: ({\bf 2},{\bf 1},{\bf 1}),\ \ \
    H_R {\rm \ and\ }\hat{H}_R: ({\bf 1},{\bf 2},{\bf 1})
\end{equation}

In this model, the scalar fields acquire the vevs, $\langle
H_R\rangle=(0,f)$ and $\langle\hat{H_R}\rangle=(0,\hat{f})$, which
break the SU(4) global symmetry as well as the gauge symmetry
$SU(2)_R\times U(1)_{B-L}$ down to $U(1)_Y$ hypercharge. Without
introducing any extra scalar fields, can we apply the mismatched
mechanism to this model to obtain a tree-level quartic coupling to
the pseudo Goldstone Higgs? The previous discussion seems to
suggest that we need to change the vev of $\hat{H}_R$ to
$\langle\hat{H_R}\rangle=(\hat{f},0)$. These new vevs would break
$U(1)_Y$ and hence, $U(1)_{EM}$. Therefore, this mechanism can not
be applied to the left-right twin Higgs model in its simplest
form. The question we would like to answer is whether there exists
a different operator or a certain assignment of charges that
achieves the same goal, while leaving $U(1)_{EM}$ unbroken.

\subsection{Quartic for the left-right model}

The  charge assignment for $H$ and $\hat{H}$ given in eq.~(\ref{Hcharge})
is unique. All other charge assignments which are
consistent with the symmetry breaking $SU(2)_L\times SU(2)_R\times
U(1)_{B-L} \rightarrow U(1)_{EM}$ and preserve the left-right
symmetry are, up to a set of field redefinitions, equivalent to
this assignment. With this charge assignment, the vevs that
preserve the hypercharge $U(1)_Y$ is the one given in the original
LRTH model: $\langle H\rangle=(0,0,0,f)$ and
$\langle\hat{H}\rangle=(0,0,0,\hat{f})$. In order to have a tree
level quartic, we add to the LRTH model the following terms that break
the global SU(4) symmetry

\begin{eqnarray}\label{quartic}
    \Delta V = \lambda(|(H_R^T\tau_2\hat{H}_R)|^2
    +|(H_L^{\dagger}\hat{H}_L)|^2).
 \end{eqnarray}

These two terms are not symmetric under the twin defined originally in the LRTH model
\begin{eqnarray}\label{twin}
    H_L &\leftrightarrow& H_R\nonumber\\
    \hat{H}_L &\leftrightarrow& \hat{H}_R,
\end{eqnarray}
where the gauge and matter fields transform as
\begin{eqnarray}\label{twin2}
A_{L\mu}^{a}T_L^a & \rightarrow & A_{R\mu}^{a}T_R^{a} \nonumber\\
A_{B-L}& \rightarrow & A_{B-L}\nonumber\\
Q_L &\rightarrow &Q_R^c,
\end{eqnarray}
in two-component Weyl notation. However, one can define an
alternative twin parity
\begin{eqnarray}\label{newtwin}
    H_L &\leftrightarrow& H_R\nonumber\\
    \hat{H}_L &\leftrightarrow& \tau_2\hat{H}_R^*,\nonumber\\
    A_{L\mu}^{a}T_L^a & \rightarrow & A_{R\mu}^{a}T_R^{a} \nonumber\\
    Q_L &\rightarrow &Q_R^c,
\end{eqnarray}
It can be shown explicitly that the quartic terms given in
eq.~(\ref{quartic}) preserve the $Z_2$ symmetry given in
eq.~(\ref{newtwin}), which is as powerful as the original twin
parity in protecting the Higgs mass from receiving quadratically
divergent corrections. All interactions in this model except the
$U(1)_{B-L}$ gauge interaction  and the new quartic
potential we introduced in eq.~(\ref{quartic}) preserve both of
the parities given above. The quartic potential breaks the first
parity and the $U(1)_{B-L}$ breaks the second.

Since every term in this extended LRTH model breaks no more than one parity defined in
eqs.~(\ref{twin},\ref{twin2}) and eq.~(\ref{newtwin}), quadratically divergent masses
of the PNGB can only be generated when both parities are broken collectively. The
quadratically divergent contributions to the PNGB masses are generally expected to
arise at two loop. However, a more detailed analysis shows that two-loop contributions
are also absent, and that contributions begin at three loops. We have thus succeeded
in constructing a tree level quartic without generating a large mass term for the
Higgs.

\subsection{SU(4) invariant top Yukawa interaction}

Since precision measurements prefer a light Higgs, $m_h< 200$ GeV
\cite{higgsbound}, a tree level quartic by itself is not as useful
as one might hope in addressing the LEP paradox. In order to have
a complete solution to the problem, a further suppression of Higgs
mass parameter is desirable. An obvious way to achieve this is to
extend the top sector to include a $U(4)$ invariant Yukawa and
terms that only break the global symmetry softly. Then, the Higgs
potential will receive only a finite contribution from the top
sector {\cite{chacko}}.

The top sector in the original LRTH model contains $Q_{L,R}$ and
$T_{L,R}$ charged under ${\rm SU}(3)_c\times {\rm SU}(2)_L\times
{\rm SU}(2)_R\times {\rm U}(1)_{B-L}$ as
\begin{eqnarray}
    Q_{L} &=& ({\bf 3},{\bf 2},{\bf 1},{\bf 1/3})\;\;\;\;\;\;
    Q_{R} =({\bf\bar{3}},{\bf 1},{\bf\bar{2}},-{\bf 1/3})\nonumber\\
    T_L &=& ({\bf 3},{\bf 1},{\bf 1},{\bf 4/3})\;\;\;\;\;\;\;
    T_R = ({\bf \bar{3}},{\bf 1},{\bf 1},{\bf -4/3}),
\end{eqnarray}
where we are using two-component Weyl notation. The gauge
invariant top Yukawa terms can then be written down as
\begin{eqnarray}\label{topyukawa}
    y(H_R^{\dagger}\tau_2{Q}_RT_L+H_L^T\tau_2{Q}_LT_R).
\end{eqnarray}
Without introducing any more extra fields, all other quarks and
charged leptons can get their masses from non-renormalizable
operators like
\begin{eqnarray}\label{topyukawa}
    \frac{y_u}{\Lambda}(H_R^{\dagger}\tau_2{Q}_R) (H_L^T\tau_2{Q}_L)
    +\frac{y_d}{\Lambda}(H_R^{T}{Q}_R) (H_L^{\dagger}{Q}_L)
\end{eqnarray}
Due to the smallness of the Yukawa couplings, these
non-renormalizable operators will not affect our discussion later
of fine tuning. Even after we have modified the Higgs sector by
adding a new quartic term eq.~ (\ref{quartic}), the charges of the
Higgses and their vevs remain the same as that were defined
originally and thus these operators remain valid to give masses to
light fermions. We will ignore these operators for the rest of
this paper.

Notice that neither $Q_{R}$ nor $Q_{R}^c$, the complex conjugate
of $Q_R$, can be combined with $Q_L$ to form an SU(4) multiplet
due to the different charges under the gauge or Lorentz groups. To
complete the SU(4) representation, we need to introduce two extra
vector-like quarks
\begin{eqnarray}
    \Phi_{R} &=&({\bf 3},{\bf 1},{\bf 2},{\bf 1/3})\;\;\;\;\;\;\;
    \Phi_{L} = ({\bf\bar{3}},{\bf 2},{\bf 1},-{\bf 1/3})\nonumber\\
    \bar{\Phi}_{R} &=&({\bf\bar{3}},{\bf 1},{\bf 2},-{\bf 1/3})\;\;\;\;
    \bar{\Phi}_{L} = ({\bf 3},{\bf 2},{\bf 1},{\bf 1/3}).\nonumber
\end{eqnarray}
$Q_L$ and $\Phi_R$ form a {\bf 4} representation of SU(4) and
similarly for $\Phi_L$ and $Q_R$. The top Yukawa term then becomes
\begin{eqnarray}
    {\cal L}_{top}&=&y(H_L^T\tau_2\Phi_L+H_R^T\tau_2{Q}_R)T_L\nonumber\\
           &+&(H_L^T\tau_2{Q}_L+H_R^T\tau_2\Phi_R)T_R + h.c. .
\end{eqnarray}
We also add the following soft masses to decouple the extra vector-like quarks,
\begin{eqnarray}
    M_R\bar{\Phi}_R\Phi_R +M_L\bar{\Phi}_L\Phi_L + M_0T_LT_R + h.c. .
\end{eqnarray}
For simplicity, we set $M_0=0$ in the analysis below.

\section{radiative corrections and EW symmetry breaking}

In this section we determine the radiative corrections to the pseudo-Goldstone mass
and verify that electroweak symmetry is indeed broken by a light Higgs. In particular,
we will compute the CW potential \cite{CW} for the light fields, as given by
\begin{eqnarray}
    V &=& \pm \sum_i \frac{1}{64\pi^2}M_i^4(\ln \frac{\Lambda^2}{M_i^2}+\frac{3}{2}),
\end{eqnarray}
where the sum is over all degrees of freedom.  The sign is positive for fermions and
negative for bosons. At one loop the Yukawa couplings, gauge couplings and Higgs
self-couplings all contribute separately to the sum, simplifying the calculation. For
simplicity, we will work in the context of a model where the symmetry breaking pattern
is realized linearly, by the terms
\begin{eqnarray}
\label{SU4higgs}
  &&\eta(|H|^2-f^2)^2+\hat{\eta}(|\hat{H}|^2-\hat{f}^2)^2.
\end{eqnarray}

We begin by considering the loop contributions from the
self-couplings of the scalar fields. Obviously, there can be no
$\eta$ or $\eta^2$ contribution to the potential of Goldstone
bosons since all vertices in the relevant diagrams preserve
$SU(4)$. Hence, these diagrams will only correct $\eta$, a free
parameter. Also, to one-loop, the diagrams with one mismatched
quartic and one SU(4) invariant quartic ($\eta\lambda$
contribution) will only generate corrections to $\eta$ and
$\lambda$, both free parameters. This can be understood by the
observation that
\begin{eqnarray}
    &&\lambda(|(H_R^T\tau_2\hat{H}_R)|^2
    +|(H_L^{\dagger}\hat{H}_L)|^2)\\
    &&=\lambda|H_L^{\dagger}\hat{H}_L
    +H_R^{\dagger}i\tau_2\hat{H}_R^*|^2+\lambda|H_L^{\dagger}\hat{H}_L-H_R^{\dagger}i\tau_2\hat{H}_R^*|^2.\nonumber
\end{eqnarray}
The first operator is invariant under an SU(4), which is also
preserved by $\eta$, if we arrange
$\hat{H}=(\hat{H}_L,i\tau_2\hat{H}_R)$. The same holds for the
second if we arrange $\hat{H}=(\hat{H}_L,-i\tau_2\hat{H}_R)$. At
one loop, the four-point diagrams that include the SU(4) invariant
quartic can only include one of these operators and thus are
invariant under the corresponding SU(4). Hence, the combination of
the operators above will only correct the tree level parameters
$\eta$ and $\lambda$. Therefore, when computing the one loop
radiative corrections to quartic terms in the Higgs potential, we
can ignore the SU(4) invariant term given in eq.~(\ref{SU4higgs}).

The effective potential may however contain operators of higher
dimensionality involving $\eta$ arising at one loop, but these
operators will make only a finite contribution to the potential of
the pseudo-Goldstone bosons. We will therefore neglect this
contribution in our analysis. As mentioned in the previous
section, new quadratic contributions could arise from the
combination of the quartics above and the $U(1)$ gauge coupling at the
three loop level, which we will also ignore.

The vev that preserves $U(1)_{EM}$ can be written as
\begin{equation}
    \langle H\rangle = f\left(%
\begin{array}{c}
  0 \\
  i\sin{x} \\
  0 \\
  \cos{x} \\
\end{array}%
\right) \;\;\;\;\;\;\langle\hat{H}\rangle = \hat{f}\left(%
\begin{array}{c}
  0 \\
  i\sin{\hat{x}}\\
  0 \\
  \cos{\hat{x}}\\
\end{array}
\right)
\end{equation}
Expanding the tree level Higgs potential given in
eq.~(\ref{quartic}) and keeping only the mass terms we find
\begin{eqnarray}
    &\lambda&\{|\hat{f}\cos\hat{x}H_{R1}-f \cos
    x\hat{H}_{R1}|^2\nonumber\\
    &+&|f\sin x\hat{H}_{L2}-\hat{f}\sin\hat{x}H_{L2}^*|^2\nonumber\\
    &+&f\hat{f}\sin x\sin\hat{x}(H_{L}^{\dagger}\hat{H}_{L}+ h.c.)\}.
\end{eqnarray}
For the right-handed fields, obviously three of them are
massless and the last one has mass squared
$\lambda(\hat{f}^2\cos^2\hat{x}+f^2 \cos^2 x)$.
For the left-handed fields, the eigenvalues are
$\pm\lambda f\hat{f}\sin x\sin\hat{x}$, $\lambda f^2 \sin^2 x$,
$\lambda \hat{f}^2 \sin^2 \hat{x}$ and
\begin{eqnarray}
&&\frac{1}{2}\lambda(\hat{f}^2\sin^2\hat{x}+f^2 \sin^2 x)\nonumber\\
&& \;\;\;\; \pm \sqrt{f^4 \sin^4 x + \hat{f}^4 \sin^4 \hat{x} + 14 \hat{f}^2 f^2 \sin^2 x \sin^2 \hat{x}}.\nonumber
\end{eqnarray}

It is now clear how the quadratically divergent mass terms for
the pseudo-Goldstone bosons vanish. The quadratic terms in the one-loop
CW potential are proportional to $\sum_i M_i^2$. From the masses
given above, the trace is not zero but independent of $x$ and
$\hat{x}$, which are the two Higgs fields.

We now turn our attention to contributions arising from the top Yukawa coupling.
The masses of fermions in the top quark sector are given by
\begin{eqnarray}
    &&\frac{1}{2}(f^2+M^2\pm\sqrt{(f^2+M^2)^2-4M^2f^2\sin^2
    x})\nonumber\\
    &&\frac{1}{2}(f^2+M^2\pm\sqrt{(f^2+M^2)^2-4M^2f^2\cos^2
    x}),
\end{eqnarray}
where we have imposed a left-right symmetry to the soft masses, so
$M_L=M_R=M$. Again, the sum of $M_i^2$ is independent of $x$.

Finally, we turn our attention to the gauge sector. The
masses of the gauge bosons are
\begin{eqnarray}
    m^2_{W_H} &=& \frac{g_2^2}{2}(f^2+\hat{f}^2)-m_W^2\nonumber\\
    m_{Z_H}^2 &\approx & \frac{g_1^2+g_2^2}{2}(f^2+\hat{f}^2)-\frac{2g_1^2+g_2^2}{g_1^2+g_2^2}m_W^2.
\end{eqnarray}

To quadratic order, the CW potential is
\begin{equation}\label{loopmass}
V_2^{(1)}=v^2(V_a+V_b\cos^2\beta),
\end{equation}
where
\begin{eqnarray}
V_a &=&\frac{1}{32\pi^2}\nonumber\\
&&\{\frac{3}{2}g_2^4(f^2+\hat{f}^2)(\ln\frac{\Lambda^2}{m_{WH}^2}+1)\nonumber\\
&&+3\frac{2g_1^2+g_2^2}{4}g_2^2(f^2+\hat{f}^2)(\ln\frac{\Lambda^2}{m_{ZH}^2}+1)\nonumber\\
&&+2\lambda^2(f^2+\hat{f}^2)(\ln\frac{\Lambda^2}{\lambda (f^2+\hat{f}^2)}+1)\}\label{lambdamass},\\
V_b &=& \frac{1}{32\pi^2}12y^2\frac{M^2}{y^2f^2-M^2}\nonumber\\
&&(y^2f^2\ln\frac{y^2f^2+M^2}{y^2f^2}-M^2\ln\frac{y^2f^2+M^2}{M^2})
\end{eqnarray}
and
\begin{eqnarray}
v\sin \beta=f\sin x,\;\;\;\;\;\; v\cos\beta=\hat{f}\sin\hat{x}.
\end{eqnarray}

To align the direction of the electro-weak symmetry breaking,
we add the following soft mass terms
\begin{eqnarray}\label{softmass}
    V^{(0)}&=& m^2H_L^{\dagger}H_L+\hat{m}^2\hat{H}_L^{\dagger}\hat{H}_L\nonumber\\
           &+&\mu^2(H^{\dagger}\hat{H}+h.c.).
\end{eqnarray}
Together with the SU(4) breaking quartic term given in
eq.~(\ref{quartic}), the tree level potential is given by
\begin{eqnarray}
    V^{(0)}&=&\lambda v^4\cos^2\beta\sin^2\beta\nonumber\\
&+&v^2(m^2\sin^2\beta+\hat{m}^2\cos^2\beta+2\mu^2\sin\beta\cos\beta)\nonumber\\
&+&2\mu^2f\hat{f}\sqrt{A}
\end{eqnarray}
where
$A=(1-\frac{v^2}{f^2}\sin^2\beta)(1-\frac{v^2}{\hat{f}^2}\cos^2\beta)$.

We now minimize the potential $V=V^{(0)}+V_2^{(1)}$ to find
$v$ and $\sin\beta$. The potential $V$ has the form
\begin{eqnarray}\label{potential}
V&=&v^2(a+b\sin^2\beta+2\mu^2\cos\beta\sin\beta)\nonumber\\
&+&\lambda v^4\cos^2\beta\sin^2\beta,
\end{eqnarray}
where
\begin{eqnarray}
    a&=&\hat{m}^2-\mu^2\frac{f}{\hat{f}}+V_a\nonumber\\
    b&=&m^2-\hat{m}^2-
    \frac{\mu^2}{f\hat{f}}(\hat{f}^2-f^2)+V_b.
\end{eqnarray}
After minimization, we find
\begin{eqnarray}
    \sin^2\beta &=& \frac{a}{2a+b}\nonumber\\
    v^2 &=& -\frac{2a+b}{\lambda}(1+\frac{\mu^2}{\sqrt{a(a+b)}}).
\end{eqnarray}

The fine tuning is about 13\% for $\hat{f}=2.0$ TeV and about 18\%
for $\hat{f}=1.6$ TeV, with the feature that $\lambda$ is much
less than $1$. Unfortunately, this is not significantly better
than the original twin model. Notice that a mass squared is
generated at loop level proportional to $\hat{f}^2\lambda^2$ (See
eq.~(\ref{lambdamass})). Since $\hat{f}$ must be greater than 1.6
TeV to evade the bound from direct $Z'$ and $W'$ gauge boson
searches \cite{pdg}, the $\hat{f}^2\lambda^2$ contribution to the
mass squared could be large if we push $\lambda$ too high, which
will tend to increase fine tuning. Thus, a small $\lambda$ is
preferred. However, with a smaller $\lambda$, we should account
for the one-loop contribution to the quartic, since it may no
longer be negligible. The largest loop contribution to the quartic
is from the top Yukawa and is given by
\begin{equation}
\label{topcontribution}
   V^{(1)}_4=\lambda_t v^4 \sin^4 \beta,
\end{equation}
where
\begin{eqnarray}
    \lambda_t&=&\frac{3}{16\pi^2}y^4\frac{M^4}{m_T^4}\left\{\ln\frac{m_T^2}{m_t^2}-\frac{1}{2}\right.\\
    &+&(\frac{m_T^2}{2M^2-m_T^2})^3\left.\ln\frac{M^2}{m_T^2-M^2}-2(\frac{m_T^2}{2M^2-m_T^2})^2\right\}\nonumber
\end{eqnarray}
and
\begin{eqnarray}
    m_T^2&=&
    M^2+y^2f^2,\;\;\;\;\;\;m_t^2=\frac{M^2}{m_T^2}y^2v^2\sin^2\beta.
\end{eqnarray}
After adding eq.~(\ref{topcontribution}) to eq.~(\ref{potential}) and repeating the
analysis above, we find that a fine tuning of about 30\% is easily achieved. Selected points are
shown in table~(\ref{little_twin_table}).
\begin{table}[tbh]
\begin{tabular}{|c|c|c|c|c|c|c|} \hline
$\Lambda${\tiny (TeV)} & $f${\tiny (GeV)} & $\hat{f}${\tiny (TeV)}
& $M_{L,R}${\tiny (TeV)} & $m_{h}${\tiny (GeV)} &$\sin^2\beta$ &
Tuning \\ \hline
10 &800 &1.6 &4 &150/233 &0.54 &0.30 ($y$)\\\hline 10 &800 &3.5
&4&150/236 &0.54 &0.10 ($\hat{f}$)\\\hline 20 &1600
&1.6&4&163/213 &0.66 &0.11 ($M$)\\\hline 10 &800 &1.6 &10&147/266
&0.51 &0.19 ($y$)\\\hline 5 &800 &1.6 &4 &150/233&0.54
&0.30 ($y$)\\\hline 5
&800 &3.5 &4 &150/236&0.54 &0.16 ($\hat{f}$)\\
\hline 10 &1600 &1.6 &4&163/213 &0.66 &0.11 ($M$)\\
\hline
\end{tabular} \caption{ A summary of the Higgs mass and fine
tuning, $\partial \; {\rm log} M_Z^2 / \partial \; {\rm log} {f}^2
$, for sample points of parameter space. The two values of $m_h$ correspond to
the masses of the two neutral Higges.  The most fine tuned
parameter at each point is shown in the parenthesis. At these points, the
other parameters are $\mu^2=-(150$ GeV$)^2$, $\lambda=0.5$ and
$y=\sqrt{2}$.}\label{little_twin_table} 
\end{table}

\section{mirror model}

As far as addressing the little hierarchy problem, the mirror twin Higgs model with a
tree level quartic \cite{nomura} provides an improvement over the original mirror
model. However, as shown in section II, in this theory the mirror photon is
necessarily massive. As a consequence, this theory has difficulty in explaining the
absence of a mirror electron relic density. In the absence of a massless mirror
photon, electrons cannot efficiently annihilate to photons. We now show that using the
same mechanism that was discussed in the previous section, this difficulty can be
avoided.

The gauge group in the mirror model is $SU(2)_A\times U(1)_A\times
SU(2)_B\times U(1)_B$ which is a subgroup of the global U(4)
symmetry. The scalar fields are $H$ and $\hat{H}$ which have the
same charge under the gauge group. The top sector is just the SM
top Yukawa plus its twin counter part.
\begin{eqnarray}
    {\cal L}_{top}&=&y(H_L^T\tau_2Q_L t_R+H_R^T\tau_2{Q}_R t_L)
\end{eqnarray}

We now calculate the CW potential in this model. The masses of
heavy gauge bosons are
\begin{eqnarray}
    m^2_{W_H} &=& \frac{g_2^2}{2}(f^2+\hat{f}^2)-m_W^2\nonumber\\
    m_{Z_H}^2 &=& \frac{g_1^2+g_2^2}{2}(f^2+\hat{f}^2)- \frac{g_1^2+g_2^2}{g_2^2}m_W^2.
\end{eqnarray}
For the top sector, up to finite terms which do not significantly
alter the fine tuning, we can just take $M=\Lambda$ to produce the
results that correspond to the non-SU(4) invariant top sector. For
the Higgs potential, we add the same tree level potential as given
in eq.~(\ref{quartic}). The CW potential due to this tree level
potential is exactly the same as that obtained in our previous
analysis on the left-right model. To quadratic order, the
potential is
\begin{eqnarray}\label{loopmass}
    V_2^{(1)}&=&\frac{v^2}{32\pi^2}\nonumber\\
    &\{&\frac{3}{2}g_2^4(f^2+\hat{f}^2)(\ln\frac{\Lambda^2}{m_{WH}^2}+1)\\
    &+&\frac{3}{4}(g_1^2+g_2^2)^2(f^2+\hat{f}^2)(\ln\frac{\Lambda^2}{m_{ZH}^2}+1)\nonumber\\
    &+&2\lambda^2 (f^2+\hat{f}^2)(\ln\frac{\Lambda^2}{\lambda (f^2+\hat{f}^2)}+1)\nonumber\\
    &-&12y^4f^2\sin^2 \beta
    (\ln\frac{\Lambda^2}{y^2f^2}+1)\}.\nonumber
\end{eqnarray}
The one-loop quartic from the top sector is
\begin{eqnarray}
    V_4^{(1)}&=&\frac{3}{16\pi^2}y^4[\ln\frac{\Lambda^2}{m_t^2}+\ln\frac{\Lambda^2}{m_T^2}+\frac{3}{2}]\nonumber
\end{eqnarray}
where
\begin{eqnarray}
    m_T^2&=&y^2f^2\;\;\;\;,\;\;\;\;m_t^2=y^2v^2\sin^2\beta.
\end{eqnarray}
We then analyze the effective potential given by
$V=V^{(0)}+V_2^{(1)}+V_4^{(1)}$ as in the previous section. The
fine tuning for this model is shown in table (\ref{mirror_table}).
We see that the results represent an improvement over the mirror
model. We expect that further enhancement may be obtained by
making the top Yukawa coupling SU(4) invariant as in
{\cite{chacko}}, but we leave this for further work.

\begin{table}[tbh]
\begin{tabular}{|c|c|c|c|c|c|} \hline
$\Lambda${\tiny (TeV)} & $f=\hat{f}${\tiny (GeV)} &$\lambda$ &
$m_{h}${\tiny (GeV)}  & Tuning \\ \hline 10 &800 &0.5 &178/213
 &0.16 ($y$)\\\hline 10 &800 &1 &183/213
&0.21 ($y$)\\\hline
\end{tabular} \caption{ A summary of the Higgs mass and fine
tuning, $\partial \; {\rm log} v^2 / \partial \; {\rm log} {f}^2
$, for sample points of parameter space. The two values of $m_h$ correspond to
the masses of the two neutral Higges. The most fine tuned
parameter at each point is shown in the parenthesis. At these points, the other
parameters are $\mu^2=-(150$ GeV$)^2$, $y=1.2$ and
$\sin^2\beta=0.69$.} \label{mirror_table}
\end{table}

\section{Conclusion}
We have constructed a twin Higgs model based on left-right
symmetry with an order one tree level quartic for the light Higgs.
The structure of the electroweak symmetry breaking is similar to
that of two Higgs doublet model. We analyzed the model and showed
electroweak symmetry breaking can happen naturally. For $\hat{f}
=1.6$ TeV, which is the lower bound from the direct searches on
heavy gauge bosons, the fine tuning is found to be about $30\%$
for $\Lambda=10$ TeV. The bound on $\hat{f}$ gets stronger if we
also require the left-right symmetry on the first two generation
quarks. The $K_0$-$\bar{K}_0$ mixing puts a very strong constraint
on the mass of $W_H$ which require $\hat{f}> 3.5$ TeV \cite{KLKS}.
In that case, the fine tuning is found to be about $10\%$.

The phenomenology of the model introduced in section II and III is
not significantly different from that of the original left-right
twin Higgs model\cite{phenoLRTH,phenoLRTH3,phenoLRTH2}. The extra
quarks introduced to complete the SU(4) multiplet could have
masses of about 4 TeV
which is difficult to observe at the LHC. 
Among these extra quarks there are some with electric charge
$Q=-1/3$. These new down-type fermions in the model might have
sizable contributions to the $D^0-\bar{D}^0$ mixing depending on
their masses\cite{D0}. The current experimental bound can be used
to put a bound on the parameter $M$ in the model. Another
difference is that the parity we introduced to make the
$\hat{h}_L$ stable, under which $\hat{H}_L$ is odd and all other
fields are even, is here softly broken by the term
$H_L^{\dagger}\hat{H}_L$ in eq. (\ref{softmass}). Hence,
$\hat{H}_L$ is no longer a dark matter candidate and will be
produced and decay just like all other scalars in the model. The
phenomenology of the scalar sector of the original LRTH model has
been studied in ref. \cite{phenoLRTH,phenoLRTH2}. Most of these
studies have focused on the scalars in the `right-handed' $H_R$
and $\hat{H}_R$ since all other scalars in the `left handed'
sector other than the SM Higgs do not interact directly with
fermions. In both of our new models, for the same reason that
$\hat{H}_L$ is no longer stable, all scalars in the `left-handed'
sector can interact with fermions and will behave just like the
scalars of two Higgs doublet model. This new phenomenon, probably
in combination with some others, may be used to test the tree
level quartic coupling introduced in these twin Higgs models. We
leave these studies for future work.

In summary we have shown how to incorporate a tree level quartic into the left-right
twin Higgs model, leading to a substantial improvement in fine-tuning.  We have
further applied this mechanism to the mirror twin Higgs model and established that the
fine tuning is about $20\%$ for a 10 TeV cutoff scale.

\noindent {\bf Acknowledgments --} We thank Zackaria Chacko for
discussions and comments on the draft. C. K and H.S.G are
supported by the NSF under grant PHY-0408954.

\end{document}